\documentclass[aps,pre,twocolumn,groupedaddress,showpacs,floatfix]{revtex4}
\usepackage{amsmath}
\usepackage{amssymb}
\usepackage{graphicx}

\begin{document}

\title{Topographical Maps as Complex Networks}
\author{Luciano da Fontoura Costa and Luis Diambra}
\affiliation{Institute of Physics of S\~ao Carlos. University of
S\~ ao Paulo, S\~{a}o Carlos, SP, Caixa Postal 369, 13560-970,
phone +55 16 3373 9858,FAX +55 162 3373 9879, Brazil,
luciano@if.sc.usp.br}

\date{13 Agosto 2004}

\begin{abstract}

The neuronal networks in the mammals cortex are characterized by
the coexistence of hierarchy, modularity, short and long range
interactions, spatial correlations, and topographical connections.
Particularly interesting, the latter type of organization implies
special demands on developing systems in order to achieve precise
maps preserving spatial adjacencies, even at the expense of
isometry. Although object of intensive biological research, the
elucidation of the main anatomic-functional purposes of the
ubiquitous topographical connections in the mammals brain remains
an ellusive issue. The present work reports on how recent results
from complex network formalism can be used to quantify and model
the effect of topographical connections between neuronal cells
over the connectivity of the network. While the topographical
mapping between two cortical modules are achieved by connecting
nearest cells from each module, four kinds of network models are
adopted for implementing intra modular connections, including
random, preferential-attachment, short-range, and long-range
networks. It is shown that, though spatially uniform and simple,
topographical connections between modules can lead to major
changes in the network properties in some specific cases,
depending on intra modular connections schemes, fostering more
effective intercommunication between the involved neuronal cells
and modules. The possible implications of such effects on cortical
operation are discussed.

\end{abstract}


\maketitle
\section{Introduction}

Among the vast spectrum of natural phenomena involving intense
information exchange between spatially distributed elements, the
mammals cortex stands out as particularly complex and intriguing
\cite{Scannell:1999}. While the full understanding of this organ still
represents one of the biggest challenges to science, continuing
investigations in the areas of neuroanatomy, neurophysiology,
neurogenetics, and neuroinformatics have provided a wealthy of
information about its organizational principles.  It is currently
known that the mammals neocortex is characterized by coexistence of
hierarchy, modularity, short and long range connections, spatial
correlations, and topographical maps.  Two neuronal modules are said
to be topopographically connected if adjacent neurons of the input
layer connect to adjacent neurons of the output layer
(e.g. \cite{Chklovskii:2000, Kandel:1995}).  One of the most
distinctive and ubiquitous properties of the mammals brain are the
topographical connections between its several modules
\cite{VanEssen:1997, Bartels_Zeki:1998, Kandel:1995}.  In the visual
system, for example, the cortical region LGN (i.e. lateral geniculate
nucleus) connects topographically to the cortical region V1, and then
to V2 and further (e.g. \cite{Ding:1998, Ichida:2002, Lyon:1998}).
The fact that such modules exchange information vertically along the
hierarchies while communicating horizontally with other modules in the
same hierarchical level has motivated a computational model known as
multistage integration \cite{Bartels_Zeki:1998}.  Indeed, cortical
feedback through such connections seems to be essential for achieving
important functionalities such as orientation selectivity
\cite{Murphy:1999}. Sensory cortical modules are not only
topographically connected between themselves, but also receive
topographically structured representations of the input stimuli.  In
the visual system, for instance, we have the visiotopic maps which
have been identified as being important for target detection
\cite{Kardar:2002}. At the same time, the functional characteristics
of the cells distributed along the cortical surface have been shown to
be highly correlated, in the sense that cells that are close one
another tend to have similar functions
(e.g. \cite{Kandel:1995,Friston:1996}).  It is very likely that such
organizational principles are not accidental or secondary.
Contrariwise, it is possible that such geometrical arrangement of the
cortical circuitry may be essential for proper information
processing. The main purpose of the present work is to analyze modular
topographical connections in terms of the interesting and powerful
concepts and measurements supplied by the new area of complex
networks. The basic structure of the cortical connections, which is
shared by many species, is likely to be the result of an attempt at
optimizing several conflicting requirements simultaneously, including
minimal wiring, minimal metabolism/energy, number of cortical areas,
as well as molecular and genetical constraints
\cite{Karbowski:2003}. The minimal wiring requirement, and henceforth
minimal transmission delay, have often been identified in the
literature as the most important requirement shaping cortical
connections
\cite{Cherniak:1995,Mitchison:1991,Karbowski:2001,Chklovskii:2002}.
The special efforts invested by the central nervous system in
implementing topographical connections provide a primary indication
that such a kind of strategy plays an important role in minimizing
connectivity requirements while guaranteeing effective cortical
processing.

The importance of maintaining spatial relationships and adjacencies
through several cortical modules can, in principle, be associated to
the following putative requirements or properties derived from
experimental findings and computational theory:

\emph{Adjacency:} As extensively indicated by experimental
investigations, neighboring neurons tend to have similar
functionalities, implying spatial correlation of neuronal activity
along the cortical surface.  Such an organization also accounts for a
certain degree of redundancy.

\emph{Accessibility:} Neural operation involves intensive exchange of
information along time and space.  In order that decisions can be
taken timely, it is important that neurons enrolled in cooperative
processing have effective access to information in any of the enrolled
cells.  Information accessibility can be quantified in terms of time
or distance, and can be estimated inside the same cortical module or
between different modules.  High accessibility demands more
connections between neurons, with the highest possible degree being
achieved when each cell is directly connected to every other cell.  In
other words, connectivity tends to favor accessibility.

\emph{Parallelism:} As neuronal cells are relatively slow processing
units, real-time cortical operation requires parallel and distributed
processing.  It is important to observe that some parallel processing
paradigms, such as vector processing and pipelining, do not
necessarily lead to intense combinatorial connections between all
involved modules.

\emph{Broadcasting:} Another important mechanism possibly underlying
information transmission is \emph{broadcasting}.  Unlike
point-to-point intercommunication, broadcasting is characterizing by
the fact that the same information is sent to several other neurons.
Broadcasting can be particularly useful for neural modulation and
control.  While wide broadcasting can take place along time, short
term action demands high levels of neuronal connections.

Though it is not currently clear how these features are adopted
and combined at different cortical regions in order to allow
emergence of proper behavior, it is only through the quantitative
characterization of network connectivity and spatial constraints
that new hypotheses and further experimental investigation,
including functional evaluation, can be obtained and validated. In
order to better appreciate the possible implications of
topographical connections for cortical architecture, it is also
important to consider the connectivity patterns intrinsic to
cortical modules. As there is no current agreement on whether the
local cortical connections follow random \cite{Braitenberg:1998}
or selective attachment \cite{Young:1992,Shefi:2002}, both
situations are considered in the present study.

While the connections underlying traditional neuronal networks can
naturally be represented in terms of graphs, the recent interest
in \emph{complex networks} \cite{Albert_Barab:2002} has paved the
way to characterize important properties of such structures with
respect to both their connections and organization, especially in
terms of the node degree (i.e. the number of connections of each
neuron), average length and clustering coefficient
\cite{Albert_Barab:2002, Bollobas:2002, Barabasi_Ravasz:1998}.
Graph theory \cite{Bollobas:2002} thus provides several concepts
and tools for measuring, modeling and validating several aspects
of cortical geometry and functionality.

The present work reports on the potential of applying complex
network formalism to investigate in a quantitative manner the
effects that topographical connections may have in modifying
important properties of the involved cortical modules, with
special attention given to their connectivity. The neuronal cells,
each represented by a network node, are assumed to be uniformly
distributed along the cortical topography. The connections inside
each modular network follow four different architectures:
preferential-attachment (PAT), random network (RAN), short-range
network (SHR) and long-range network (LNR), which will be detailed
later. Topographical connections between two such cortical modules
A and B are established by linking each node of A to the nearest
node in B with probability $\beta$. In order to quantify the
impact of such topographical connections over the network
properties, {several measurements are obtained while the degree of
topographical coupling between modules A and B, quantified in
terms of the probability $\beta$, is increased}. Furthermore, we
measure the same network properties for random connections between
A and B in order to establish a comparison with the topographical
connections case. We also consider uni- and bidirectional
connections between modules for both topographical and random
inter-modular models, as further explained below.

The article starts by introducing the adopted network terminology
as well as the several measurements considered for
characterization of the properties of the investigated networks.
The obtained simulation results characterize the changes of the
network properties in terms of the topographical coupling,
indicating that topographical connections can have major impact
over the properties of the involved networks.

\section{Models and methods}

\begin{figure}
\begin{center}
\hspace*{-0.5cm}
\includegraphics[angle=0,scale=0.75]{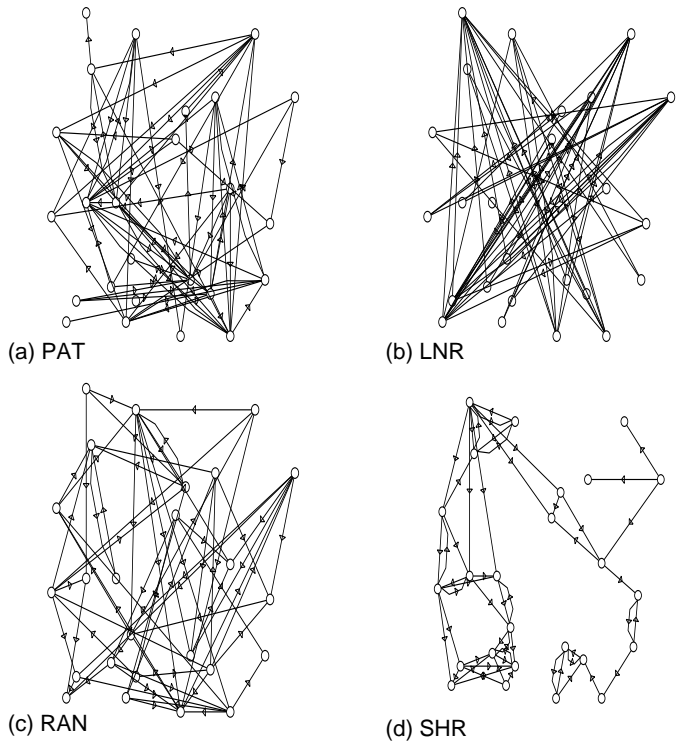}
\end{center}
\caption{Diagram representing the four typical network modules
used in this work: preferential-attachment (A), long range (B),
random (C) and short range (D). The spatial distribution of nodes
is the same for all diagrams. We considered $L=50$, $p=0.1$,
$\gamma= 0.01$ (i.e. $N=25$ nodes), and $T=0.03$ for the cases B
and D.~\label{fig:redes}}
\end{figure}

The cortical modules, with $N$ nodes connected through $n$
directed edges, are embedded into an $L \times L$ two-dimensional
space $\Omega$ representing the cortical domain associated to each
module. Each network node $i$, $i=1,2, \ldots, N$, is randomly
positioned at coordinate $(x_i, y_i)$. The number of nodes in each
module is $N ={\rm Round}\left(\gamma L^2 \right)$ where $\gamma$
is the density of nodes and \rm{Round} is the rounding function.
The connections inside each cortical module are implemented as
follows: a pair of selected nodes establishes a connection if a
random number, uniformly distributed between [0,1], is smaller
than $p$. The selection of nodes to be connected is performed
according to the following architectures: PAT, RAN, SHR or LNR. In
the PAT networks, the connections follow a preferential attachment
scheme as described in the following. The probability to choose a
node to implement a connection depends linearly on the number of
connections of that node. Our procedure began with the same
probability for all nodes and new connections are added
progressively. This procedure results in a network with few highly
connected nodes and many poorly connected nodes (Figure
~\ref{fig:redes}-a). For RAN networks, two nodes are selected at
random for respective connection (Figure ~\ref{fig:redes}-c). The
construction of the SHR and LNR networks are similar but follow
different criteria.  First, a list of pair of nodes with Euclidean
distances in decreasing or increasing order is elaborated (for SHR
or LNR networks, respectively).  Then a pair of nodes is randomly
selected and if a random number, uniformly distributed between
[0,1], is smaller than $\exp\left[-i/(TN)\right]$, the pair is
connected, where $i$ is the ordinal number of the pair in the list
and $T$ is a dimensionless parameter that regulates the length of
the connection. Small values of $T$ produce networks that observe
the imposed criteria more closely, while for very large values of
$T$ a RAN architecture is obtained (see Figures ~\ref{fig:redes}-b
and ~\ref{fig:redes}-d). In this paper we set $T=0.1$ in order to
retain some longer or shorter range intra-modular connections for
SHR and LNR networks, respectively. For all cases the boundary
conditions of modules are open (i.e. non-periodical).

\begin{figure}
\begin{center}
\includegraphics[angle=0,scale=0.5]{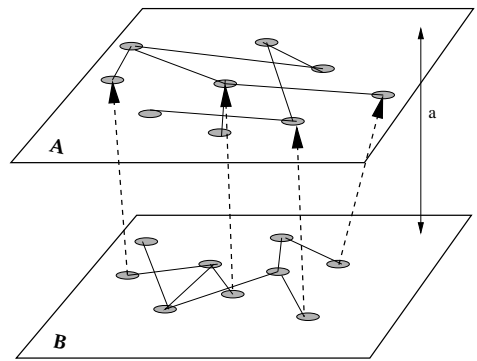}
\end{center}
\caption{Diagram representing unidirectional topographical
connections (dashed lines) from nodes of module B to nodes of
module A, which are separated by distance $a$.~\label{fig:rede2}}
\end{figure}

Random and topographical schemes were considered for the
inter-modular connections (IMC). In the topographical case, the
IMC between neurons in the two cortical modules A and B are
established according to the following rule: each neuron in module
A connects in a directed way to the nearest neuron in module B
with probability $\beta$. The nearest neighbor condition is
ignored in the case of the random model. Therefore, the
probability $\beta$ controls the degree of topographical coupling
between the two modules, with complete mapping being achieved for
$\beta=1$. The two modules A and B are adjacent and they are
separated by a distance $a$, so that nodes in B do not overlap
with nodes in A (see Figures ~\ref{fig:rede2} and
~\ref{fig:rede1}). Both topographical and random IMC models were
implementing regarding bidirectional or unidirectional
connections. In the bidirectional case, there exist directed
connections from nodes of module A to nodes of module B, and
vice-versa from B to A, while for unidirectional connections, there
exist directed connections solely from nodes of module B to nodes
in module A.

The following measurements were used in this work in order to
quantify several properties of the considered complex networks:

\emph{Shortest path length:} Let $i$ and $j$ be two network nodes with
at least one path from $i$ to $j$. The \emph{shortest path},
$\delta_{i,j}$, between these nodes is understood as the minimal total
sum of edge lengths connecting $i$ to $j$. The average of this
measurement, denoted by $\left< \delta \right>$, can be used to
characterize the accessibility between two nodes in terms of Euclidean
arc length. When there are no paths from $i$ to $j$, we set
$\delta_{i,j}= \left(N-1 \right) \sqrt{ \left( 2 L^2+a^2 \right)}$ in
order to penalize the accessibility in that case.  We calculate the
mean over all shortest paths of the network in the bidirectional case,
while for unidirectional case the averaging procedure is applied only
over $\delta_{i,j}$ such that $i$ and $j$ are nodes from B and A
layers respectively, as illustrated in Figure ~\ref{fig:rede2}.

\emph{Path degree:} Is the fraction of pair of nodes without any
path between them. Represented henceforth as $\left< \pi \right>$,
this measurement provides complementary information about the
accessibility between nodes.

\begin{figure}
\begin{center}
\hspace*{-0.25cm}
\includegraphics[angle=0,scale=0.7]{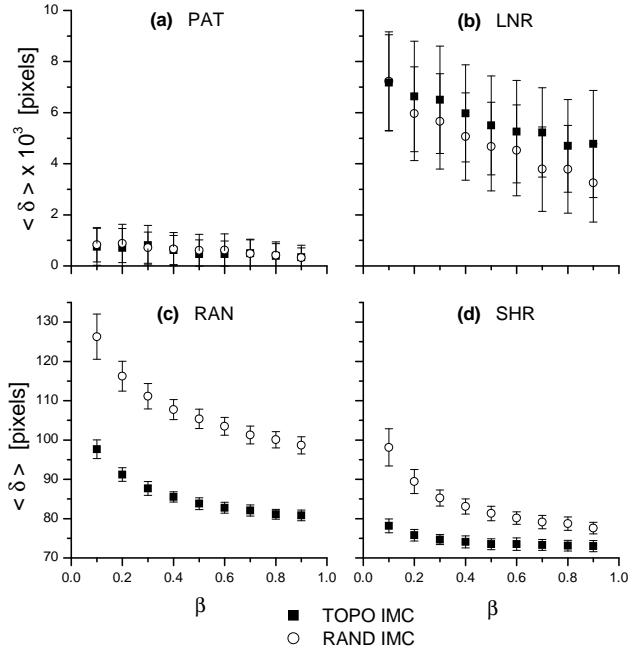}
\end{center}
\vspace*{-0.75cm}

\caption{The average and standard deviation of the shortest path
lengths for unidirectional IMC networks with $p=0.1$. We consider four
different modular architectures: preferential-attachment (a), long
range (b), random (c) and short range (d). Filled squares correspond
to topographic IMC and open circles to random IMC. The scale of the
vertical axes in (b) and (d) panels are the same as in (a) and (c)
respectively.  ~\label{fig:dic1}}
\end{figure}

\begin{figure}
\begin{center}
\hspace*{-0.25cm}
\includegraphics[angle=0,scale=0.7]{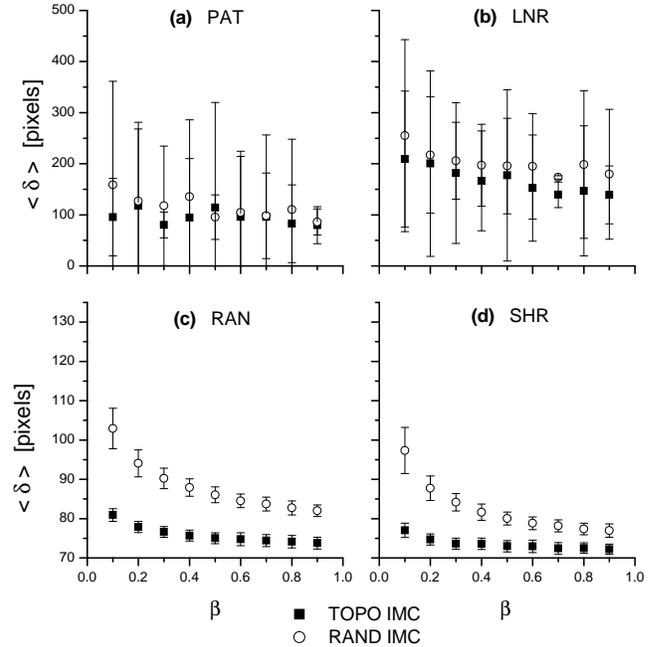}
\end{center}
\vspace*{-0.75cm} \caption{The average and standard deviation of
shortest path lengths for unidirectional IMC networks with
$p=0.3$. We consider four different modular architectures:
preferential-attachment (a), long range (b), random (c) and short
range (d). Fill square corresponds to topographic IMC and open
circle to random IMC. The scale of the vertical axes of (b) and
(d) panels are the same that (a) and (c) respectively.
~\label{fig:dic3}}
\end{figure}

\section{Results}

One hundred realizations of each models were obtained by
simulations considering $L=128$ pixels, $\gamma=0.02$, $p=0.1$ and
$0.3$. For short range networks, we used $T=0.1$, while the values
of $\beta$ vary from 0.1 to 0.9. The adopted cortical separation
was $a=8$ pixels, similar to the typical distance between
neighboring nodes inside a module.

\begin{figure}
\begin{center}
\includegraphics[angle=0,scale=0.5]{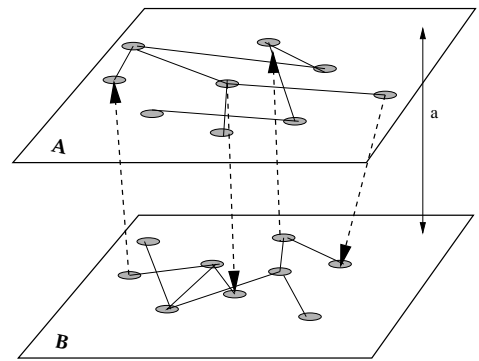}
\end{center}
\caption{Diagram representing two network modules (A and B)
separated by distance $a$ and connected bidirectionally by
topographical connections (dashed lines).~\label{fig:rede1}}
\end{figure}

\begin{figure}
\begin{center}
\hspace*{-0.25cm}
\includegraphics[angle=0,scale=0.7]{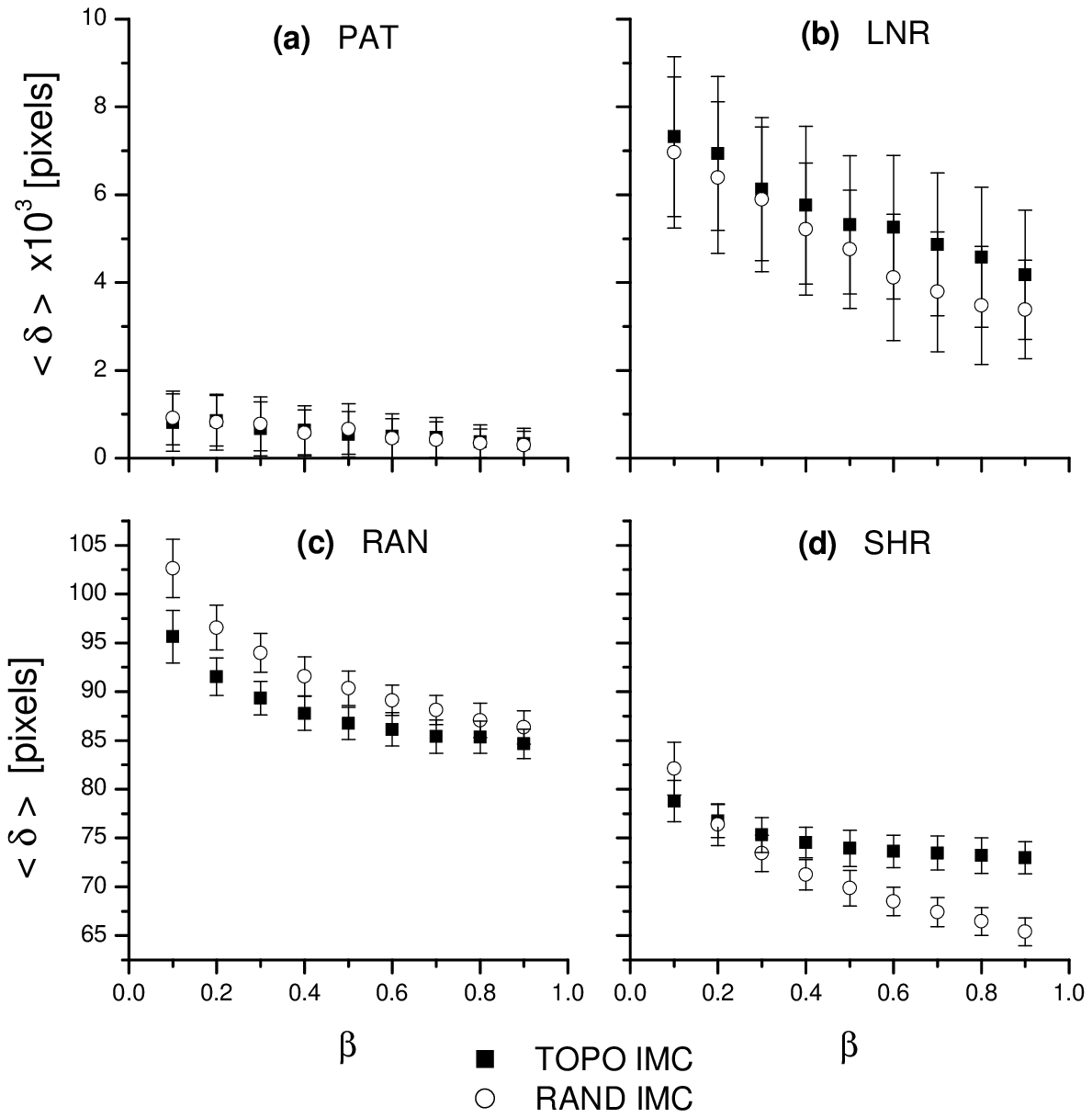}
\end{center}
\vspace*{-0.75cm} \caption{The average and standard deviation of
shortest path lengths for bidirectional IMC networks with $p=0.1$.
We consider four different modular architectures:
preferential-attachment (a), long range (b), random (c) and short
range (d). Fill square corresponds to topographic IMC and open
circle to random IMC. The scale of the vertical axes of (b) and
(d) panels are the same that (a) and (c) respectively.
~\label{fig:bidic1}}
\end{figure}

\begin{figure}
\begin{center}
\hspace*{-0.25cm}
\includegraphics[angle=0,scale=0.7]{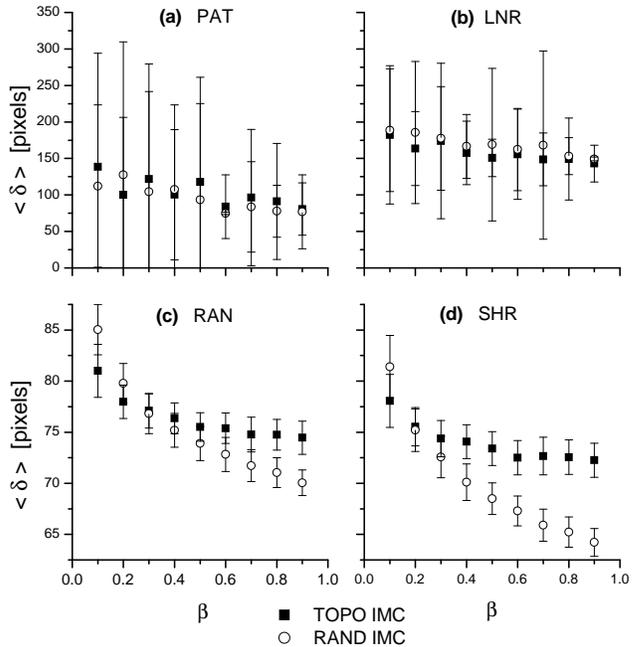}
\end{center}
\vspace*{-0.75cm}

\caption{The average and standard deviation of the shortest path
lengths for bidirectional IMC networks with $p=0.3$.  We consider four
different modular architectures: preferential-attachment (a), long
range (b), random (c) and short range (d). Filled squares correspond
to topographic IMC and open circles to random IMC. The scale of the
vertical axes in (b) and (d) panels are the same as in (a) and (c)
respectively.  ~\label{fig:bidic3}}
\end{figure}

Figures~\ref{fig:dic1} and~\ref{fig:dic3} show the average
shortest path length $\left< \delta \right>$ as a function of the
unidirectional IMC intensity quantified by $\beta$, for $p=0.1$
and $p=0.3$, respectively. The values $\left< \delta \right>$
decrease linearly in the PAT and LNR architectures
(Figures~\ref{fig:dic1}(a-b) and~\ref{fig:dic3}(a-b)), while the
RAN and SHR architectures present an exponential decay of $\left<
\delta \right>$ when $\beta$ increases
(Figures~\ref{fig:dic1}(c-d) and~\ref{fig:dic3}(c-d)). The IMC
type, either topographical or random, influences the behavior of
$\left< \delta \right>$ in a selective manner, depending on the
connectivity model adopted for the modules. For example, in PAT
and LNR architectures the behavior of $\left< \delta \right>$ for
random IMC is essentially undistinguishable from the topographical
IMC case. The greatest difference appears for RAN and SHR
architectures: in these cases the topographical connections have a
lower average shortest path than the corresponding random IMC. The
difference between topographical and random IMC is more evident in
the RAN architecture for $p=0.1$. In addition, when $p$ increases
from 0.1 to 0.3, $\left< \delta \right>$ decreases by about 20$\%$
in the RAN architecture for both topographical and random IMCs,
while in the SHR architecture $\left< \delta \right>$ does not
present significant changes. This is in contrast to the strong
decrease in $\left< \delta \right>$ when $p$ increases from 0.1 to
0.3 in the PAT and LNR architectures, particularly in the latter
case, which always showed $\left< \delta \right>$ higher than the
former.

Figures~\ref{fig:bidic1} and~\ref{fig:bidic3} display the average
shortest path length, $\left< \delta \right>$ as a function of the
IMC intensity $\beta$, for $p=0.1$ and $p=0.3$ respectively,
regarding bidirectional connections. We observe that the IMC type,
either topographical or random, influences the behavior of $\left<
\delta \right>$ in a selective manner, depending on the modular
architecture. In particular, $\left< \delta \right>$ associated to
random IMC is lower than that corresponding to topographical IMC
only for the SHR architecture. In the case of  the other modular
architectures there are no great differences. In a similar way to
the unidirectional case, $\left< \delta \right>$ decreases
linearly in the PAT and LNR architectures
(Figures~\ref{fig:bidic1}(a-b) and~\ref{fig:bidic3}(a-b)), and
exponentially in the RAN and SHR architectures
(Figures~\ref{fig:bidic1}(c-d) and~\ref{fig:bidic3}(c-d)). In
addition, we observe that topographical IMC decays slower than
random IMC. The difference between the two types of IMC is more
definite in the SHR architecture for $p=0.3$. We also note that an
increase of $p$ values from 0.1 to 0.3 in the PAT and LNR
architectures, particularly in the latter, tend to enhance
accessibility between the network nodes in that models
(Figures~\ref{fig:bidic1}(a-b) and~\ref{fig:bidic3}(a-b)). This
effect is substantially weaker in the RAN architecture
(particularly in the topographical IMC), and almost null in the
SHR architecture both for topographical and random IMCs.

\begin{figure}
\begin{center}
%\hspace*{-0.25cm}
\includegraphics[angle=0,scale=0.4]{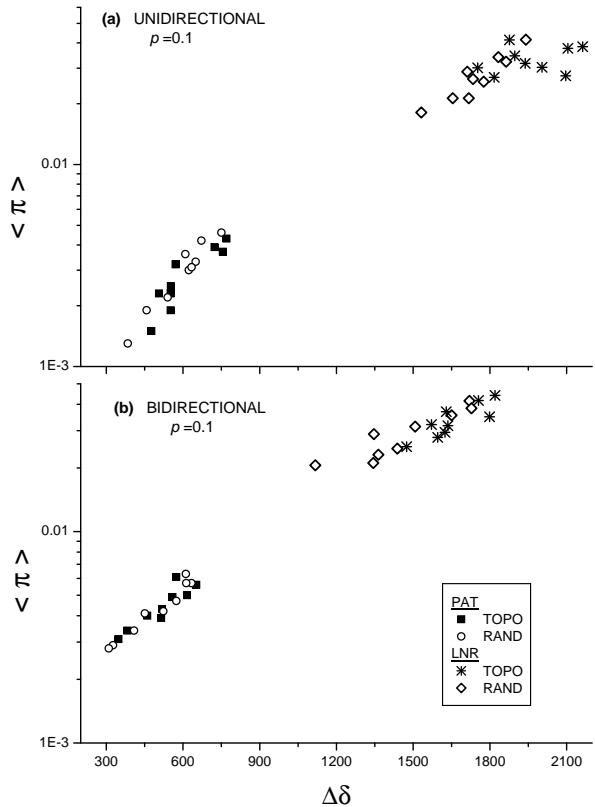}
\end{center}
\vspace*{-0.75cm} \caption{Fraction of pairs of nodes without a
path connecting them, $\left< \pi \right>$, {\it versus} the
standard deviation of the shortest path lengths, $\Delta \delta$,
for unidirectional (a) and bidirectional (b) connections types. We
consider PAT networks with topographical IMC (filled squares), and
random IMC (open circles); and LNR networks with topographical IMC
(asterisk symbols), and random IMC (open
rhomboids).~\label{fig:cor}}
\end{figure}

The properties characterized by the node degree, path degree and
clustering coefficient did not show significant differences between
topographic and random realizations of the IMC. However, we observed
different values of path degree produced by the PAT and LNR when
compared to the other two network types. In particular, path degrees
tend to be considerably higher in the PAT and LNR than in SHR and RAN
networks, which implies a higher standard deviation $\Delta \delta$
(displayed as errors bars in
Figures~\ref{fig:dic1}-\ref{fig:bidic3}). Figure~\ref{fig:cor}
displays the scatter plots of $\left< \pi \right>$ in terms of the
standard deviation $\Delta \delta$ of the shortest path length. We
observe that the fraction of pair of nodes without any path connecting
them is smaller in the LNR networks than in PAT networks. Moreover
$\left< \pi \right>$ suffers the influence of directionality (uni- or
bidirectional) in the PAT networks but not in LNR case. In the latter
case, random IMC seems to be more efficient than topographical both
for unidirectional and bidirectional cases. The measurements $\left<
\pi \right>$ for RAN and SHR networks are small and almost independent
of the IMC type, $p$ and $\beta$ values (for the values considered
here), while in PAT networks $\left< \pi \right>$ decreases with
$\beta$ and $p$.

Our simulations suggest that topographical networks are more
effective in terms of minimal wiring only for the RAN and SHR
architectures with unidirectional IMCs. However, opposite
conclusions were reached regarding bidirectional IMCs, except for
the RAN architecture for low $p$ and $\beta$. Furthermore, the
cortical architectures obtained for the short range networks are,
in general, more effective in the sense of minimal wiring than
random networks, and less sensitive to the IMC intensity.

\section{Conclusions}

An interesting conclusion from our experiments is that the PAT and LNR
models, irrespective of $p$ or the directionality of the mappings, are
little affected by the type of map between modules being random or
topographical. In other words, in case the intracortical connections
are PAT or LNR, there is little advantage in using topographical maps
as the means for getting overall shorter connections. In that case,
topographical maps would need to be biologically justified in some
other way, perhaps in terms of metabolical constraints.  Still
regarding the PAT and LNR models, the average shortest path has been
found to be always smaller for the former, but such an advantage tends
to diminish with $p$. Therefore, preferential-attachment networks such
as those considered in this work are particularly efficient for
obtaining shortest path connections between two cortical modules
irrespective of the type of connection (i.e. random or topographical).

A completely different scenario has been identified for the RAN and
SHR models, in the sense that the type of connection between modules
tended to influence more definitely the respective average shortest
paths. Generally, the SHR tended to have average shortest path $\left<
\delta \right>$ smaller than for RAN networks.  Substantial
differences of $\left< \delta \right>$ as consequence of random or
topographical connections were observed for SHR model with
bidirectional maps and high $p$, with the random connections leading
to smaller shortest path values than the topographical case (see
Figures-\ref{fig:bidic1}(d) and-\ref{fig:bidic3}(d)).  However, the
greatest differences of $\left< \delta \right>$ were obtained for the
RAN networks with unidirectional maps and for smaller $p$ (see
Figures-\ref{fig:dic1}(c) and-\ref{fig:dic3}(c)).  Interestingly, the
topographical connections allowed substantially shortest paths in this
case. This effect was also verified, to a lesser degree, for the RAN
with higher $p$.  From the biological perspective, such results
indicate in the case of SHR and RAN cortical modules, the
topographical maps lead to substantially smaller values of average
shortest paths when one considers unidirectional maps. In this sense,
the existence of unidirectional topographical connections in the
cortex could be understood as being compatible with random and/or
short range intracortical connections, which are the connectivity
schemes found to benefit the most from such a kind of mapping.

The main implications for cortical function and organization of the
Finding reported in the current work are discussed in the
following. First, it is clear that topographical connections, even at
moderate levels, can affect the properties of the involved
modules. Indeed, the addition of a few short length connections
between the two topographically organized modules can considerably
enhance the accessibility between any two nodes in the resulting
structure, reducing the shortest path between pairs of neurons, with
the consequent improvement of time accessibility.  Such enhancements
imply that information can be exchanged and broadcasted more
effectively between the neurons of the resulting topographically
connected network than in the cortical modules taken isolated.

The intrinsic properties of topographical connections suggest that
the many cortical regions involving such a communication could
therefore account for one of the explanations for this ubiquitous
and peculiar aspect of cortical architecture, enhancing
accessibility while minimizing the length of the IMC, while also
preserving spatial adjacency. More general conclusions considering
the whole cortex are precluded by the fact that the cortical
morphology and mapping seem to vary from species to species and
from region to region (e.g.  \cite{Lyon:1998} and
\cite{Kaas:2001}).  At the same time, the use of the concepts and
results reported in this paper provides an interesting tool for
investigating and interpreting each of such cases.  Considering
that topographical connections coexist with lateral connections at
intra-modular level, the above proposed methodology can also be
immediately extended to investigate the possible implications of
such connections along the cortical modules inside the same
module.  The consideration of topographical connections between
other types of spatially-constrained complex networks can also be
considered as a means of enhancing information exchange.

\begin{acknowledgments}
Luciano da F. Costa is grateful to FAPESP (processes 99/12765-2 and
96/05497-3), CNPq and Human Frontier for financial support. Luis
Diambra thanks Human Frontier for his post-doc grant.
\end{acknowledgments}

\bibliography{cortcom}
\end{document}